# Wound Tissue Segmentation in Diabetic Foot Ulcer Images Using Deep Learning: A Pilot Study

Mrinal Kanti Dhar[1], Chuanbo Wang[1], Yash Patel[1], Taiyu Zhang[1], Jeffrey Niezgoda[2], Sandeep Gopalakrishnan[2,3], Keke Chen[4], and Zeyun Yu[1,*]

[1]Big Data Analytics and Visualization Lab, Department of Computer Science, University of Wisconsin-Milwaukee, Milwaukee, WI, USA
[2]Auxillium Health, Milwaukee, WI, USA
[3]Wound Healing and Tissue Repair Analytics Laboratory, School of Biomedical Sciences & Health Care Administration, University of Wisconsin Milwaukee, Milwaukee, WI, USA
[4]Department of Computer Science, Marquette University, Milwaukee, WI, USA

mdhar@uwm.edu, chuanbo@uwm.edu, yspatel@uwm.edu, taiyu@uwm.edu, jniezgoda@webcme.net, sandeep@uwm.edu, keke.chen@marquette.edu, and yuz@uwm.edu

*Corresponding author

## Abstract

Identifying individual tissues, so-called tissue segmentation, in diabetic foot ulcer (DFU) images is a challenging task and little work has been published, largely due to the limited availability of a clinical image dataset. To address this gap, we have created a DFUTissue dataset for the research community to evaluate wound tissue segmentation algorithms. The dataset contains 110 images with tissues labeled by wound experts and 600 unlabeled images. Additionally, we conducted a pilot study on segmenting wound characteristics including fibrin, granulation, and callus using deep learning. Due to the limited amount of annotated data, our framework consists of both supervised learning (SL) and semi-supervised learning (SSL) phases. In the SL phase, we propose a hybrid model featuring a Mix Transformer (MiT-b3) in the encoder and a CNN in the decoder, enhanced by the integration of a parallel spatial and channel squeeze-and-excitation (P-scSE) module known for its efficacy in improving boundary accuracy. The SSL phase employs a pseudo-labeling-based approach, iteratively identifying and incorporating valuable unlabeled images to enhance overall segmentation performance. Comparative evaluations with state-of-the-art methods are conducted for both SL and SSL phases. The SL achieves a Dice Similarity Coefficient (DSC) of 84.89%, which has been improved to 87.64% in the SSL phase. Furthermore, the results are benchmarked against two widely used SSL approaches—Generative Adversarial Networks and Cross-Consistency Training. Additionally, our hybrid model outperforms the state-of-the-art methods with a 92.99% DSC in performing binary segmentation of DFU wound areas when tested on the Chronic Wound dataset. Codes and data are available at https://github.com/uwm-bigdata/DFUTissueSegNet.

# 1 Introduction

Chronic wounds are those that do not heal in an orderly fashion, typically within three months [1]. Diabetic foot ulcers (DFUs), a prevalent type of chronic wound, are common in diabetic patients due to neuropathy caused by glycosylation and high blood sugar levels. Neuropathy reduces sensation and the ability to feel pain in the foot and is a leading cause of ulcers ranging from superficial to deep. Obesity and diabetes increase the risk of chronic wounds. Diabetes affects millions worldwide, with an expected 25% increase by 2030 [2]. DFU patients have a 30% risk of developing a foot ulcer, which can precede lower limb amputation in up to 85% of cases [3]. Chronic wounds impact patients' quality of life and can lead to amputations and death if not treated properly [4]. In the US, about 15% of Medicare beneficiaries have chronic wounds, with estimated Medicare expenditures between $28.1 and $31.7 billion annually. Diabetic wound infections are a significant contributor to these costs [5].

Wound measurements are the standard of care for managing chronic wounds [6]. Segmenting wound areas and wound tissues in DFUs helps assess the wound's severity, monitor progress over time, strategize future treatments, enables timely intervention, and facilitates clinical decision-making. Traditionally, identifying and segmenting distinct tissues relies on the trained eyes of specialists, a method prone to subjectivity and limited by the level of education and experience. Furthermore, the COVID-19 pandemic in 2020 significantly impacted healthcare, including wound care [7]. Deep learning, with its ability to learn complex patterns from vast data, disrupts this landscape.

Deep learning plays a crucial role in the segmentation of various organs within the human body, including the brain, eyes, abdomen, chest, feet, teeth, and heart [8]. The landscape of deep learning in multi-class segmentation is constantly evolving. U-Net [9], a popular CNN architecture, is specifically designed for medical image segmentation tasks. Researchers are also exploring advanced techniques like fully convolutional networks (FCNs) [10] and attention mechanisms [11] to further improve performance and robustness. The heart of these methods lies in training these networks on large datasets of labeled medical images. Each pixel is meticulously assigned to a specific tissue class, allowing the network to learn the intricate relationships between image features and tissue morphology. As the network processes more data, its ability to accurately segment unseen images improves, leading to remarkable results.

However, challenges remain, specifically in tasks such as segmenting wound tissues in DFUs, due to data scarcity and variability. This is because labeling individual tissues is far more time-consuming and complex than labeling the wound area and requires highly trained wound specialists. To the best of our knowledge, there is no publicly available DFU wound tissue dataset. Hence, we release a DFU wound tissue dataset called DFUTissue dataset for the research community to evaluate the effectiveness of wound segmentation algorithms in academic settings. Additionally, we conduct a pilot study on segmenting DFU tissues such as fibrin, granulation, and callus using deep learning. Due to the limited amount of annotated data in the DFUTissue dataset, we focus on a semi-supervised learning (SSL) setup to leverage unlabeled data. We propose a hybrid model that combines transformers and convolutional neural networks trained in an SSL fashion, which is found to be effective even for a small amount of unlabeled data compared to the existing SSL approaches [12] [13] [14]. Furthermore, we evaluate the performance of our proposed hybrid model in segmenting wound areas on the Chronic Wound dataset [15].

## 2  Literature Review

Diabetic foot ulcer segmentation can be categorized in two ways – wound area segmentation and wound tissue segmentation. While we have good literature available on wound area segmentation, very little data exists on wound tissue segmentation. The main reason for this deficiency is that creating a wound tissue dataset is a very tedious task. When it comes to DFU tissue segmentation alone, the literature becomes even narrower. Therefore, for the literature review, we did not limit ourselves to only DFUsy but rather considered other wound tissues as well.

**Wound Area Segmentation:** Here we briefly discuss wound area segmentation. Liu et al. [16] proposed WoundSeg, utilizing MobileNet and VGG16 architectures, achieving a Dice accuracy of 91.6% on their dataset of 950 images. However, they employed a watershed method for semi-automatic annotation instead of expert annotations. Wang et al. [15] applied a lightweight MobileNetv2 model to a chronic wound dataset with 810 training DFU images and 200 test images. They included a post-processing step to fill gaps in abnormal tissue and eliminate small regions, achieving a Dice score of 90.47%. Cao et al. [17] and Huang et al. [18] both utilized region proposal-based approaches. The former employed Mask R-CNN, while the latter used Fast R-CNN with GrabCut and SURF algorithms for DFU wound segmentation. However, the GrabCut algorithm, with GMM data integration and iterative minimization, produces less precise marked contours due to random information in practical applications. Mahbod et al. [19] and Dhar et al. [20] participated in FUSeg Challenge 2021 [21]. The former ensembled LinkNet and U-Net, however, the segmentation performance deteriorated when there was no wound or a very small wound region. The latter proposed a parallel spatial and channel squeeze-and-excitation module in an Efficient-B7-based encoder-decoder architecture. Kendrick et al. [22] used the DFUC2022 dataset for DFU segmentation, the largest available, with 2000 training and 2000 testing images. Their network, based on FCN32 with a modified VGG backbone, achieved a dice score of 74.47%. Yi et al. [23] and Hassib et al. [24] also used the DFUC2022 dataset. Yi et al. [23] proposed OCRNet with a ConvNeXt backbone and a boundary loss function, reaching a Dice score of 72.80%. Hassib et al. [24] applied SegFormer MiT-B5, achieving a Dice score of 69.89%. Attempts were undertaken to improve segmentation performance using ensemble methods combining SegFormer and DeepLabV3+, but no enhancement was noted.

**Wound Tissue Segmentation:** Some early attempts include traditional machine learning and image processing techniques, such as clustering, morphological operations, support vector machines, etc. Di Cataldo et al. [25] proposed an unsupervised approach for segmenting immunohistochemical (IHC) tissue images. Their method utilized unsupervised color clustering to automatically identify cancerous areas and disregard stroma. They also developed a technique based on local intensity distribution analysis to separate nuclei from the background, mitigating noise and staining issues. The watershed algorithm was used to segregate tissue clusters, and a post-processing step was introduced to merge oversplit nuclei based on chromatic characteristics, improving segmentation accuracy. They achieved an accuracy of 88.8% compared to 79.66% by supervised Support Vector Machine (SVM) based approaches. However, the cancerous tumor areas were already cropped manually from the real-life tissue images. Mukherjee et al. [26] developed an automated framework to classify chronic wound tissues (granulation, necrotic, and slough) by combining K-means clustering and texture analysis with the Support Vector Machine (SVM). They used 767 tissue regions where they first isolated the wound using a Mean-Shift algorithm, then extracted color (CIE Lab) and texture features to capture both hue and surface variations. K-means clustering segmented the wound into color-based clusters, refined by additional texture features. SVM then classifies each cluster as a tissue type. The framework achieved 87.61% overall accuracy. However, the use of pre-extracted tissue patches and the emphasis on classification over segmentation could limit its applicability. Li et al. [27] conducted a pilot study on wound tissue segmentation using K-means clustering, texture analysis, and Support Vector Machines (SVM) on clinical pressure injury photographs. Their method utilized a dataset of clinical photographs and employed thresholding, morphological operations, and texture analysis to differentiate healthy skin, slough, granulation, epithelial, and eschar tissues. The method achieved accuracies ranging from 80 to 88% for the four wound tissues. They used 64 images for training and 32 for validation. However, the study's reliability was limited by the selection of only a few subzones for evaluation from the initially generated significant number of subzones.

Considering the rapid advancements in deep learning techniques, upcoming studies should compare these modern methods with conventional machine learning approaches to ascertain their effectiveness in segmenting wound tissues. Although deep learning has been studied in different medical sectors, very few publications are available on deep learning-based wound tissue segmentation. Ramachandram et al. [28] proposed an attention-embedded encoder-decoder-based two-stage network for wound tissue segmentation. In the first stage, they segmented the wound region using an encoder-decoder network. Then they cropped the region with an appropriate bounding box and resized it to a higher resolution. The wound segmentation model is trained on 467,000 images. It is then passed to another encoder-decoder network to segment 4 wound tissue types – epithelial, granulation, slough, and eschar. The tissue segmentation model is trained on 17,000 images and evaluated on 383 images. It is reported that their Swift Medical Wound Data Set is the largest among other datasets so far. However, their model showed relatively poor performance for epithelial tissue. Furthermore, while they achieved decent results for wound area segmentation (mIoU 86.44%), the same cannot be said for tissue segmentation (mIoU 71.92%). Pholberdee et al. [29] proposed a method combining CNNs with traditional image processing for wound tissue segmentation. Using Medetec database images [30], they manually traced wound boundaries to create ground truth data. CNNs classified pixels, and morphological operations further analyzed them. A morphological close operation followed by dilation reconstructed the wound region, selecting the largest connected component as the wound region. They employed color data augmentation to augment the dataset. The model trained on 180 images and validated on 27 achieved accuracies of 72%, 40%, and 53% for granulation, necrosis, and slough tissue types. However, reliance on morphological operations and selecting only the largest connected component may overlook multiple wound regions. Sarp et al. [31] used a conditional generative adversarial network (GAN) for wound tissue segmentation, with data from eKare Inc [32] and included chronic wound tissue types (necrotic, slough, and granulation). Constructing datasets from 100 to 4000 images, they tested on fixed sets of 100 images. Their cGAN model had two networks and four loss functions, using a U-Net architecture. They found that a minimum of 2000 images was necessary for effective GAN training as smaller datasets struggled to represent data distributions or led to model overfitting, particularly those below 500 images. Despite using high-quality images, their Dice score of 90% aligned closely with the results of U-Net, Mask RCNN, and MobileNetV2. Also, their practice of a 97.6:2.4 training-to-test image ratio is not typical in deep learning. Lien et al. [33] focused on segmenting granulation tissue only, not the entire wound region. They used 219 images from 100 patients, dividing each into 32×32 patches. ResNet18 classified each patch into granulation, non-granulation, and non-wound categories, achieving a 60% Intersection over Union (IoU) score.

In this study, our contribution can be summarized as follows –

1. We release DFUTissue dataset that contains 110 annotated images containing granulation, fibrin, and callus. To the best of our knowledge, this is the first dataset on DFU wound tissues released for the research community.
2. We conduct a pilot study on segmenting wound tissues in DFUs. We propose a hybrid model trained in semi-supervised learning (SSL) fashion combining both transformers and convolutional neural networks (CNN).
3. We explore challenges associated with wound tissue segmentation and perform a pilot study to address them. In addition, we carry out extensive experiments and evaluate our model against state-of-the-art generative adversarial networks (GAN) [12] [13] and cross-consistency-based [14] models. Our model outperforms them by achieving a dice score (DSC) of 87.64%.
4. We also evaluate our hybrid model for binary segmentation of wound region on the Chronic Wound dataset [15]. We achieve a DSC of 92.99% outperforming state-of-the-art methods.

## 3 Materials

**DFUTissue dataset:** We established the DFUTissue dataset in collaboration with the Advancing the Zenith of Healthcare (AZH) Wound Center in Milwaukee, WI, to segment granulation, fibrin, and callus tissues in DFUs. It has 110 annotated foot ulcer images consisting of 78 images for training, 16 for validation, and 16 for inference, maintaining a train, validation, and test ratio of 70:15:15. Images are collected from the AZH Wound Center in Milwaukee, WI. The images are initially annotated by the Big Data Analytics and Visualization Lab – UWM and then refined and finalized by the AZH wound specialists. These images are divided into eight categories: granulation, callus, fibrin, necrotic, eschar, neodermis, tendon, and dressing. Specialists in wound care maintained dry necrotic tissues during data collection, as they act as a protective layer, but removed damp necrotic tissues indicating bacterial presence. Annotating wound tissues is challenging due to less defined boundaries between different tissues compared to distinct wound boundaries. Moreover, tissues like granulation exhibit irregular shapes during the healing process, further complicating the annotation procedure. **Table 1** shows the appearance count of each type of tissue. Given the limited number of images for the last five tissues, in this paper, we only focus on granulation, callus, and fibrin. All images have a fixed resolution of 256 × 256 obtained by zero padding. **Figure 1** displays samples from the DFUTissue dataset, where red represents fibrin, green indicates granulation, and blue signifies callus. Additionally, a set of 600 unlabeled images is provided by the same healthcare center, which we use to leverage our semi-supervised learning-based training process.

Table 1 The number of occurrences for each category of tissue in the 110 annotated images.

| Granulation | Callus | Fibrin | Necrotic | Eschar | Neodermis | Tendon | Dressing |
|---|---|---|---|---|---|---|---|
| 93 | 86 | 74 | 24 | 11 | 11 | 2 | 2 |

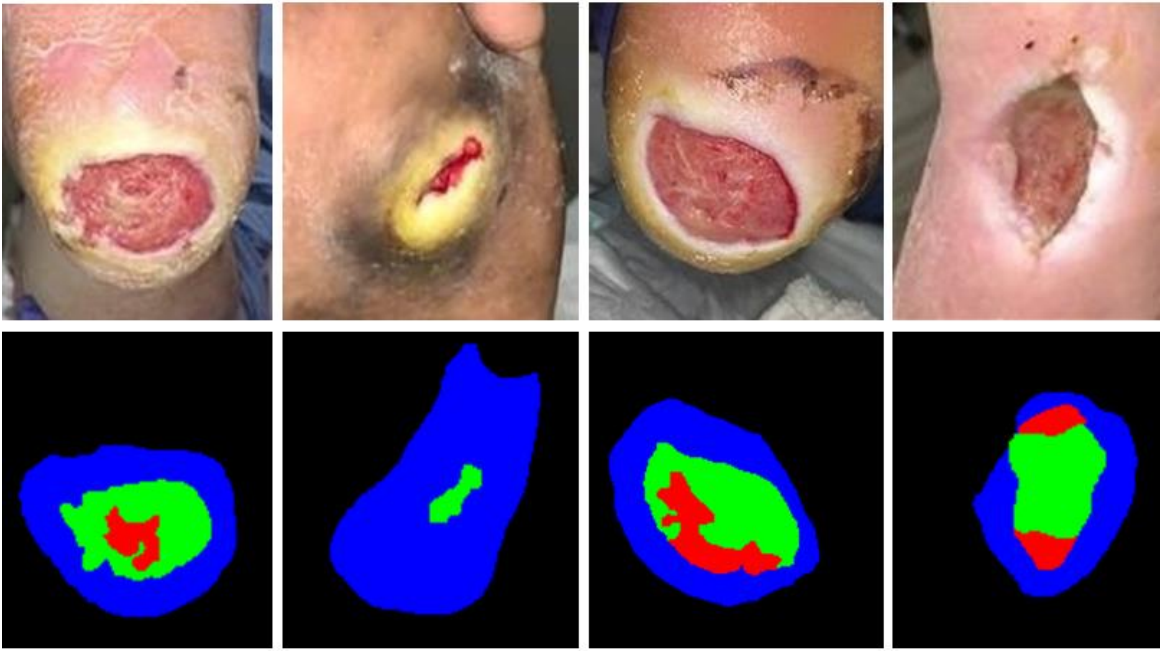

**Figure 1** Diabetic foot ulcer samples. (*top*) Chronic Wound dataset, (*bottom*) DFUTissue dataset where red, green, and blue represent fibrin, granulation, and callus, respectively. Images are cropped for better visualization.

**Table 2** Tissue distribution in the wound tissue dataset

|  | **Fibrin** | **Granulation** | **Callus** |
|---|---|---|---|
| **Images** | 74 | 93 | 86 |
| **%Images** | 67.27 | 84.55 | 78.18 |
| **Pixels** | 64929 | 166346 | 315089 |
| **%Pixels** | 2.59 | 6.63 | 12.55 |

**Tissue Characteristics:** Granulation tissue plays a vital role in the wound healing process. It demonstrates the development of angiogenesis and is an important indicator of healing. Healthy granulation tissue is characterized by its pink or light red color, moist texture, and bumpy appearance. In contrast, unhealthy granulation tissue appears pale or dark red. Fibrin is a stringy, whitish-yellow protein that forms a mesh-like structure rich with white blood cells. It's a natural part of the body's response to injury. Additionally, fibrin serves as a natural barrier against infection. Callus tissue appears as a thickened and hardened area of skin. The color of callus varies depending on factors such as skin tone, but it often tends to be yellowish or similar in hue to the surrounding skin. Callus forms in response to repeated friction, pressure, or irritation, serving as a protective response to prevent further damage to the skin.

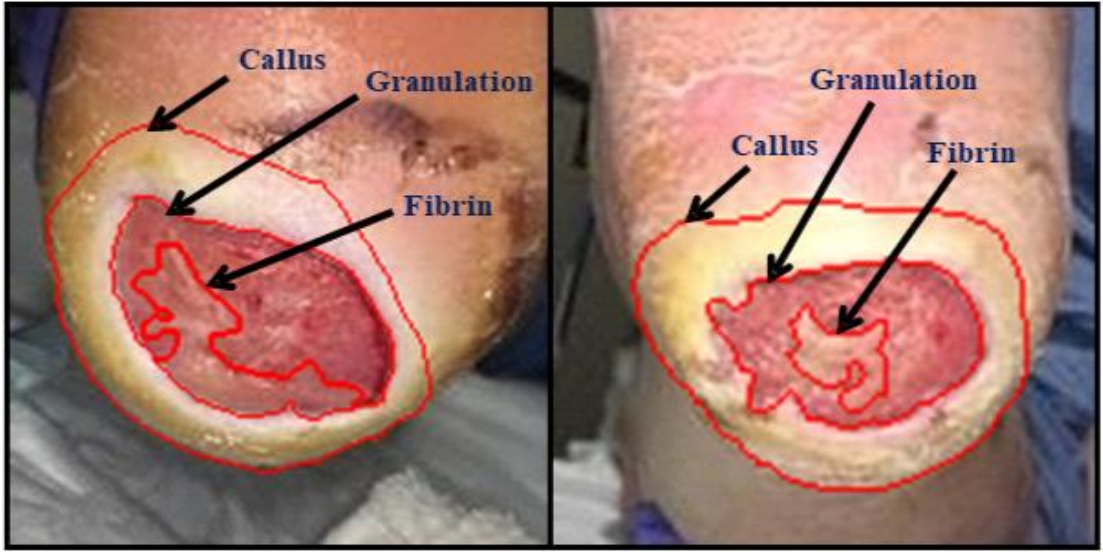

**Figure 2** Ambiguous tissue borders.

## 4 Design Strategy

Before designing the model, we first analyzed the challenges associated with this project. We found four challenges that needed to be considered.

**Challenge 1 – Very limited labeled data:** The first hurdle on our path is the scarcity of labeled data. With only 110 labeled images where 94 are available for training and validation, the model's capacity to generalize across the diverse spectrum of foot ulcer images is severely constrained. The limited dataset size raises concerns about the model's ability to capture the intricacies and variations within wound tissues, potentially hindering its overall effectiveness.

**Challenge 2 – Unbalanced tissue distribution:** As shown in **Table 2**, among 110 images, fibrin is present in 74 images, granulation in 93 images, and callus in 86 images. Hence, fibrin, accounting for 67.27% of the images, lags behind granulation (84.55%) and callus (78.18%) tissues significantly. This imbalance introduces a risk of biasing the segmentation model towards the majority classes. Furthermore, a detailed examination of pixel counts unveils the spatial prevalence of each tissue type, with granulation and callus exhibiting a considerable number of pixels, while fibrin struggles with a lower count of only 2.59% of the total pixels.

**Challenge 3 – Ambiguous tissue boundaries:** A significant impediment arises from the ambiguity surrounding tissue boundaries, particularly for fibrin and callus tissues. As shown in **Figure 2**, while granulation has relatively well-defined borders across callus tissue, distinguishing between fibrin and granulation becomes a complex task due to their ambiguous boundaries. Also, fibrin tissue resides inside the granulation area with a very similar appearance. Similarly, the demarcation between the callus and skin color poses a challenge, making it difficult to establish a clear boundary between them.

**Challenge 4 – Lack of domain-specific transfer learning:** Another notable limitation is the absence of exact domain-specific transfer learning in the project. This is particularly beneficial when working with a limited dataset, as the model can capitalize on the knowledge acquired from the larger dataset, improving its ability to generalize to new, unseen data. However, we did not find any publicly available pre-trained model specifically trained on foot ulcer images or relevant medical images.

To tackle these challenges, we adopt multiple strategies. For challenge 1, we employ a combination of both supervised and semi-supervised learning techniques. Leveraging a set of unlabeled foot ulcer data, we utilize semi-supervised learning. Given the limited amount of labeled data, L2 regularization is applied to prevent overfitting in the supervised model. Additionally, a set of data augmentation techniques is carefully designed, with different probability levels assigned to each augmentation technique. To tackle challenge 2, a hybrid loss function is designed. Moreover, we intentionally inject more augmented data containing fibrin and callus. In response to challenge 3, squeeze-and-excitation-based attention modules are added to enhance boundary tracing. As for challenge 4, despite the absence of domain-specific transfer learning, we opt for cross-domain transfer learning. The intention behind this choice is to provide the model with a head start in understanding fundamental features common across various image domains.

## 5  Methods

Our proposed framework has two phases: supervised learning (SL) and semi-supervised learning (SSL). In the SL phase, the hybrid model is trained with annotated DFU images in a supervised learning fashion. The trained model then generates pseudo-labels from the unlabeled data. Subsequently, the model enters the SSL phase for further fine-tuning. For clarity, we denote the model after the SL phase as the SL model and the model after the SSL phase as the SSL model.

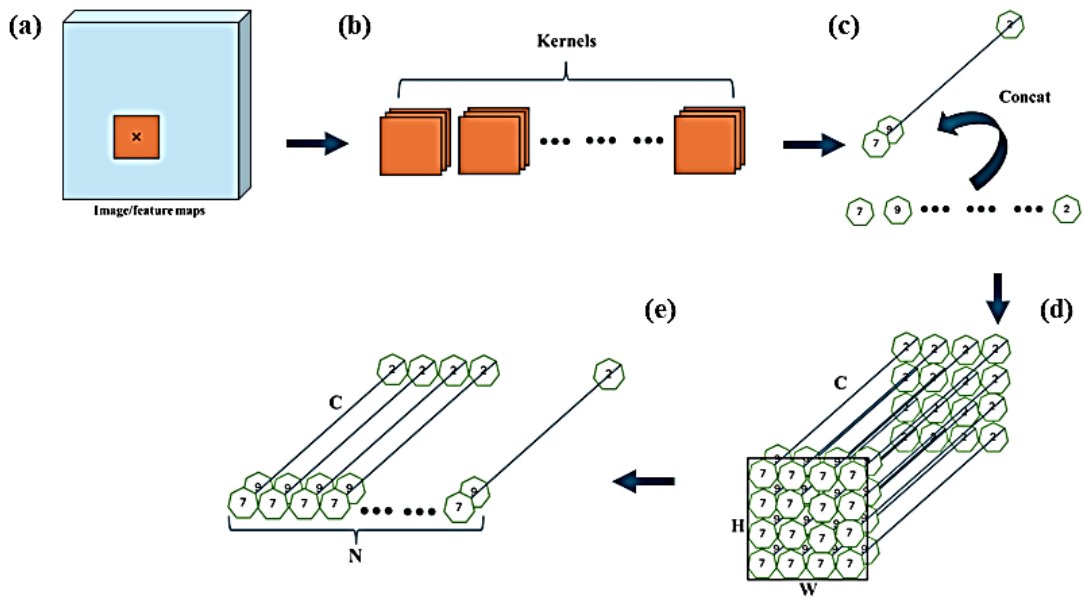

**Figure 3** Overlap patch embeddings. (a) Image or feature maps, (b) Kernels that will be applied to feature maps, (c) convolution results at pixel x obtained by the kernels. Results are stacked, (d) After performing convolution on all pixels, (e) Height and width dimensions are flattened to obtain the N×C patch embeddings.

## 5.1 Hybrid model

**Why hybrid model?** Our proposed model is a hybrid model consisting of both transformers and convolutional neural networks (CNN). The encoder-decoder architecture includes transformers in the encoder and CNN in the decoder. In addition, we integrate parallel spatial and channel squeeze-and-excitation (P-scSE) modules [20], [34] in each decoder stage. We choose to build a hybrid model so that we can combine transformer's global attention and CNN's local features. Transformers excel in capturing global dependencies within the data due to their self-attention mechanism. CNNs, on the other hand, are strong at extracting local features. Placing CNNs in the decoder allows the model to leverage localized features to create detailed segmentations. P-scSE is added due to its boundary tracing capabilities [20].

**Encoder:** In the encoder, we use Mix Transformer (MiT), a hierarchical transformer encoder initially employed in SegFormer [35]. It has four transformer blocks. Each block has three important parts – overlap patch embeddings, efficient self-attention, and Mix-FFN.

*Overlap patch embeddings:* In this stage, the model creates small patches called tokens from input images. As shown in **Figure 3**, a 2D convolution followed by flattening across spatial dimensions is applied to achieve a patching embedding of size $N \times C$, where $N$ is the sequence length and $C$ is the embedding dimension. To have the overlapped patch embedding, we need to set the stride less than the kernel size.

*Efficient self-attention:* Traditional self-attention has a computational complexity of $O(N^2)$ as the length of the sequence is $N$ [11]. Efficient self-attention reduces the length of the sequence. To do so, it reshapes the $N \times C$ matrix to $N/R \times (C \cdot R)$, where $R$ is the reduction ratio. Then it applies a linear layer to get back the embedding dimension $C$. So, the sequence length changes from $N$ to $N/R$. Therefore, the complexity reduces from $O(N^2)$ to $O(N^2/R)$.

*Mix-FFN:* In the final step, to replace positional encoding, MiT incorporates a 3×3 depth-wise convolution in the feed-forward network (FFN) with zero padding. Because studies show that for segmentation, the convolution effectively learns the positional information, eliminating the need for explicit PE [36]. Mix-FFN can be expressed as:

$$x_{out} = MLP(GELU(Conv_{3\times3}(MLP(x_{in})))) + x_{in} \qquad (1)$$

Where $x_{in}$ is the output of the efficient self-attention module.

**Decoder:** The decoder in the proposed model serves as an up-sampling path, tasked with restoring spatial information lost during the encoding process. To achieve this, essential high-resolution (albeit low semantic) information is transferred from the encoder to the decoder through shortcut connections between the two paths. As illustrated in **Figure 4**, at each stage of the decoder, the upsampled output from the lower level is initially concatenated with the encoder output from the corresponding level. The resulting concatenated output then undergoes processing through the Parallel spatial and channel squeeze-and-excitation (P-scSE) attention module, which effectively aggregates spatial and channel-wise information. This processed output subsequently undergoes a 3×3 Convolution-ReLU-Batch normalization. The P-scSE module, introduced in [20], represents an enhancement of the squeeze-and-excitation module [37], designed to boost a network's representational power by highlighting significant features and downplaying less relevant ones. As shown in **Figure 5**, the original module generates a channel descriptor through global average pooling, primarily affecting channel-related dependencies, thus being termed cSE for its channel-wise excitation. Another variant introduced by Roy et al. [38] is the sSE module, which squeezes along the channel axis and excites along spatial dimensions. The scSE module is a combination of the cSE and sSE components. The P-scSE module creates two parallel branches of the scSE module. One branch involves adding cSE and sSE, while the other branch takes the maximum of these components. The max-out mechanism introduces competition between channel and spatial excitations. The addition operation, on the other hand, aggregates these two excitations. A switch is provided to skip max-out when the number of channels is small. This design choice is grounded in the understanding that, with a limited number of channels, the model's capacity to learn intricate channel dependencies and patterns is already constrained. Thus, selectively collecting features, as max-out does, could lead to the loss of important features without contributing significantly.

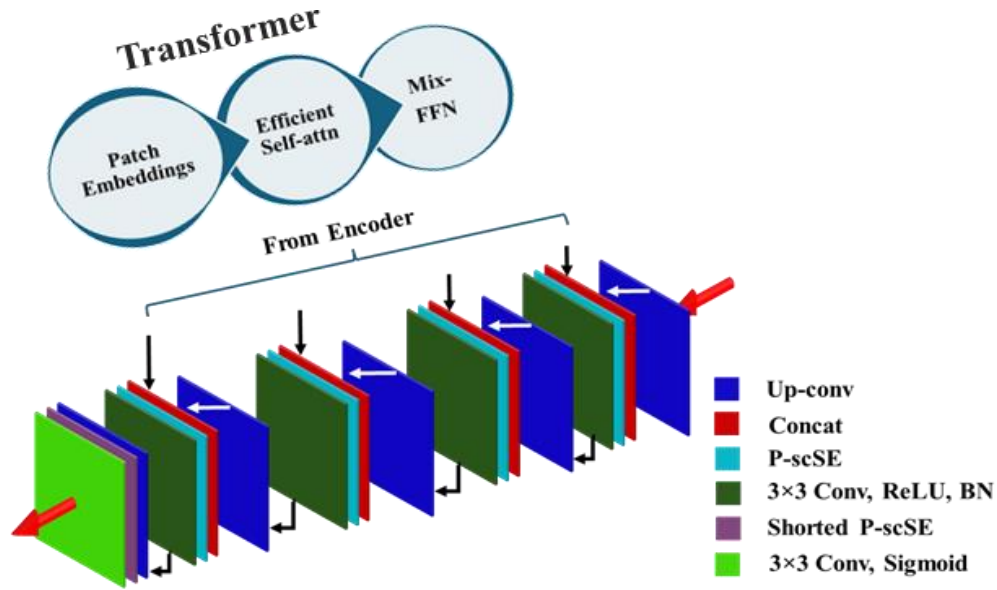

**Figure 4** Decoder structure.

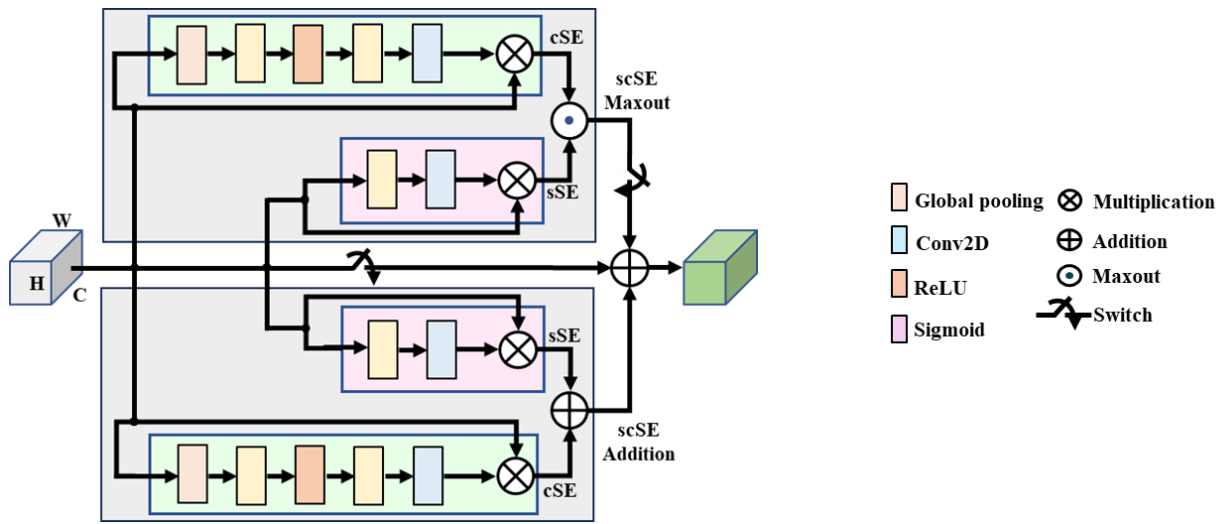

**Figure 5** Parallel spatial and channel squeeze-and-excitation (P-scSE) module.

**Algorithm 1** SemiSupervisedTraining

**Input:**
    $E$: Number of rounds (outer iterations)
    $K$: Number of runs (inner iterations)
    $U$: Pool of unlabeled images
    $L$: Pool of labeled images
    $SL$: Pre-trained supervised model
    $TV$: Track validation loss, initialize with $\infty$
    $n$: No. of images to pick randomly, default value is 50
    $VL$, $R$, $T_1$, and $T_2$: Empty lists

**Output:**
    Trained semi-supervised model

**Procedure:**
1: **For** $i \leftarrow 1$ to $E$, **do**
    a: Predict labels of unlabeled images in $U$ using the trained supervised model (SL). Store these labels in $T_1$.
    b: **For** $j \leftarrow 1$ to $K$, **do**
        i: From $T_1$, randomly pick $n$ labels and transfer them to $T_2$. Also, transfer their corresponding images from $U$ to $T_2$.
        ii: Append names of these $n$ images to $R$ as a list.
        iii: Train the supervised model ($SL$) with $T_2$. However, this time, train the model using semi-supervised loss. Keep track of the validation loss.
        iv: Append the best validation loss to $VL$.
        v: Empty $T_2$.
    c: **End for**
    d: Find the minimum validation loss (mVL) and its index from $VL$.
    e: Restore the names of this index from $R$.
    f: Transfer the images corresponding to these names from $U$ to $L$. Also, transfer corresponding labels from $T_1$ to $L$. So, $L$ now has $|L| + n$, and $U$ has $|U| - n$.
    h: **If** $TV > mVL$:
        vi: $TV \leftarrow mVL$:
    i: **Else**:
        vii: **Return** trained model
    j: Empty $T_1$, $VL$ and $R$.
2: **End for**
3: **Return** trained model

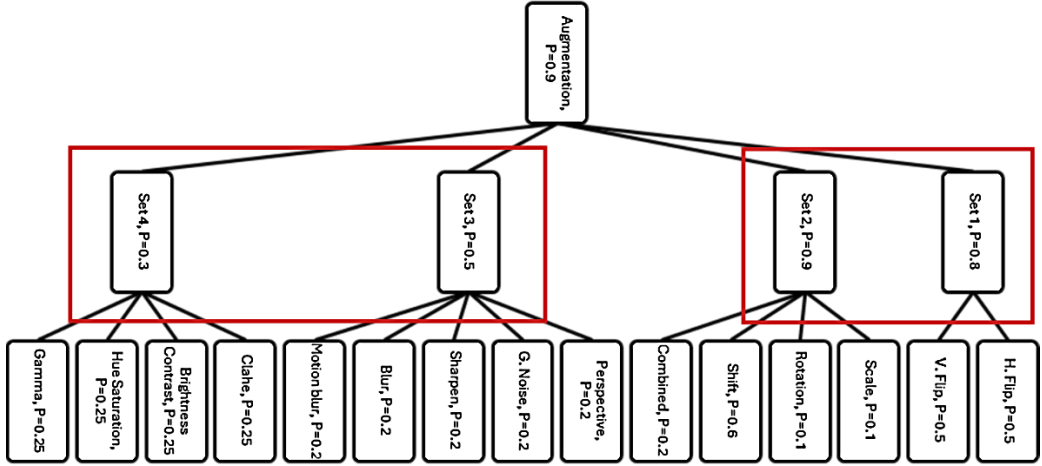

**Figure 6** Augmentation with a probabilistic approach.

## 5.2 Semi-supervised Learning (SSL)

In addition to the 110 annotated data, we have 600 more unlabeled images. Using these resources, we conduct our semi-supervised learning. Let's assume that the hybrid model is already trained with annotated data and refer to it as the SL model. As shown in **Algorithm 1**, the proposed "*SemiSupervisedTraining*" algorithm is a semi-supervised learning strategy designed to refine a recently trained SL model through iterative interactions with labeled and unlabeled image datasets. Operating over a specified number of outer iterations ($E$) and inner iterations ($K$), the algorithm begins by predicting labels for unlabeled images using the initial supervised model and storing these predictions in a temporary pool ($T_1$). In each inner iteration, a subset of $n$ labels and their corresponding images is randomly selected from the pool of unlabeled images ($U$) and transferred to another temporary pool ($T_2$). The SL model is then trained on the combination of labeled images ($L$) and the augmented pool $T_2$, incorporating a semi-supervised loss function. The algorithm keeps track of the validation loss during this training. The images contributing to the best validation loss are subsequently transferred from the unlabeled pool ($U$) to the labeled pool ($L$), effectively expanding the labeled dataset. The algorithm dynamically adjusts the labeled and unlabeled pools based on the validation loss, prioritizing informative instances. This iterative process continues until the specified number of outer iterations is completed, ultimately resulting in a refined semi-supervised model or SSL model.

## 5.3 Augmentation With a Probabilistic Approach

Yuan et al. [39] employed strong augmentation in their semi-supervised semantic segmentation, utilizing 16 different types of image transformation methods. They also included standard techniques like random scale, random crop, random flip, and normalization. The authors reported that such strong augmentation significantly enhances segmentation performance. Inspired by [39], [40], [41], we also adopt augmentation in our pipeline but with a probabilistic approach. As shown in **Figure 6**, we employ 15 image transformation methods, primarily focusing on affine and photometric transformations. Affine transformations help the model handle spatial differences in foot ulcer images, such as variations in foot position, angle, or size. Photometric transformations address differences in image appearance, like lighting conditions, contrast, and color. These augmentations generalize the segmentation model to unseen variations in spatial and photometric dimensions. The 15 methods are divided into four sets. Sets 1 and 2 pertain to affine transforms, where we prioritize flipping and shifting for affine transforms, using fewer scaling and rotations to minimize truncation and rounding errors. Except for perspective transformations, sets 3 and 4 belong to photometric transformation. Each transformation is assigned a probability for selection. This way, we can control the selection of transformations. We utilize the *Albumentations*[1] package for augmentation, adhering to its probability rules. Notably, if set 1 is selected, subsequent selections from set 2 apply transformations to an image already modified by set 1, ensuring diverse but controlled augmentation.

## 5.4 Loss Function

In this paper, we employ two distinct loss functions: one for the supervised learning phase ($L_{sl}$) and another for the semi-supervised learning phase ($L_{ssl}$). The supervised learning loss ($L_{sl}$) comprises both dice loss (DL) and focal loss (FL). Cross-entropy loss has the drawback of discretely computing per-pixel loss without taking into account whether or not the surrounding pixels are ground truth pixels, thereby ignoring the global scenario. Dice loss, originating from Sørensen–Dice coefficient, on the other hand, considers information loss both locally and globally. Dice loss can be expressed as $DL = (1 - DSC)$, where $DSC$ is the dice coefficient. Focal loss (FL) comes in handy when there is a class imbalance (for instance, background >> foreground) [42]. It down-weights easy examples and focuses training on hard (misclassified) examples or false negatives using a modulating factor, $(1 - p_t)^\gamma$, and can be expressed as:

$$FL(p_t) = -\alpha_t(1 - p_t)^\gamma \log(p_t) \quad (2)$$

Where, $\gamma > 1$ is the focusing parameter, and $\alpha_t \in [0,1]$ is a weighting factor.

For the semi-supervised phase, in addition to the dice loss and focal loss, we incorporate dynamic cross entropy (DCE) [39]. DCE is employed to address the noise present in the pseudo-labels. It is expressed as:

$$DCE_i = w_i \cdot y_i \log(\hat{y}_i) + (1 - w_i) \cdot \hat{y}_i \log(y_i) \quad (3)$$

$$w_i = max\left(\frac{exp(\hat{y}_{i0})}{\sum_{j=0}^{c} exp(\hat{y}_{ij})}, \dots, \frac{exp(\hat{y}_{ic})}{\sum_{j=0}^{c} exp(\hat{y}_{ij})}\right) \quad (4)$$

As indicated in equation (3) and equation (4), where $i$ is the location index, $\hat{y}_i$ represents the pixel prediction, $y_i$ signifies the ground truth label, and $w_i$ denotes the dynamic weight, being the maximum activation after the *softmax* operation across all $c$ classes. Prediction is employed to automatically adapt the confidence levels throughout the training process. Hence, our supervised and semi-supervised loss can be expressed as –

$$L_{sl} = DL + FL \quad (5)$$

$$L_{ssl} = \lambda_1 DL + \lambda_2 FL + \lambda_2 DCE \quad (6)$$

Where, $\lambda_1$, $\lambda_2$, and $\lambda_3$ are tuning parameters. In this paper, they are set to 1.

---

[1] https://albumentations.ai/

## 5.5 Training and Inferences

For hyperparameter tuning, we perform a grid search among weight decay values of $1\times10^{-2}$, $1\times10^{-3}$, $1\times10^{-4}$, and $1\times10^{-5}$, learning rate schedulers implemented with ReduceLROnPlateau and Poly, and optimizers such as Adam [43] and SGD. The most effective combination, resulting in the minimum validation loss, is found with a weight decay of 1×10-5, coupled with a learning rate scheduler using ReduceLROnPlateau, and employing the Adam optimizer. All experiments are executed on an NVIDIA Tesla V100 GPU provided by Google Colab Pro+, with a capacity ranging from 16 GB to 32 GB depending on availability. With a batch size of 16, we train our supervised model for 500 epochs while monitoring the validation loss and intersection-over-union (IoU) score. We keep storing and overwriting the checkpoint whenever the validation loss decreases or the IoU score increases. Therefore, only the best checkpoint is evaluated during inference. To avoid needless training, an early stopping with patience 50 is utilized.

## 5.6 Evaluation Metric

For the segmentation task, we use widely used Intersection-over-Union (IoU), Precision, Recall, and Dice Similarity Coefficient (DSC). Here are each definition's details –

$$Precision = \frac{TP}{TP + FP} \quad (7)$$

$$Recall = \frac{TP}{TP + FN} \quad (8)$$

$$DSC = \frac{2TP}{2TP + FP + FN} \quad (9)$$

$$IoU = \frac{TP}{TP + FP + FN} \quad (10)$$

Here *TP*, *FP*, and *FN* are true positive, false positive, and false negative, respectively.

## 6 Results and Discussion

### 6.1 Baseline

In this study, we analyze both the fully supervised stage, relying on annotated data, and the semi-supervised stage (SSL), which leverages both annotated and unlabeled data. According to [13], U-net [9] is reported as the most widely used segmentation network in the medical field among fully supervised methods. Additionally, DeepLabV2 [44] was employed for the same purpose in the referenced study. However, in our work, we opt for DeepLabV3+ [45] over DeepLabV2 due to its superior baseline performance. Furthermore, we include SegFormer-b3 in our analysis due to its relevance to our architecture.

For semi-supervised learning (SSL), we explore two popular streams: GAN-based and cross-consistency-based approaches. We modify a GAN model proposed by [12]. In the case of cross-consistency, we use cross-consistency training (CCT) as proposed by [14], and their codes are publicly available.

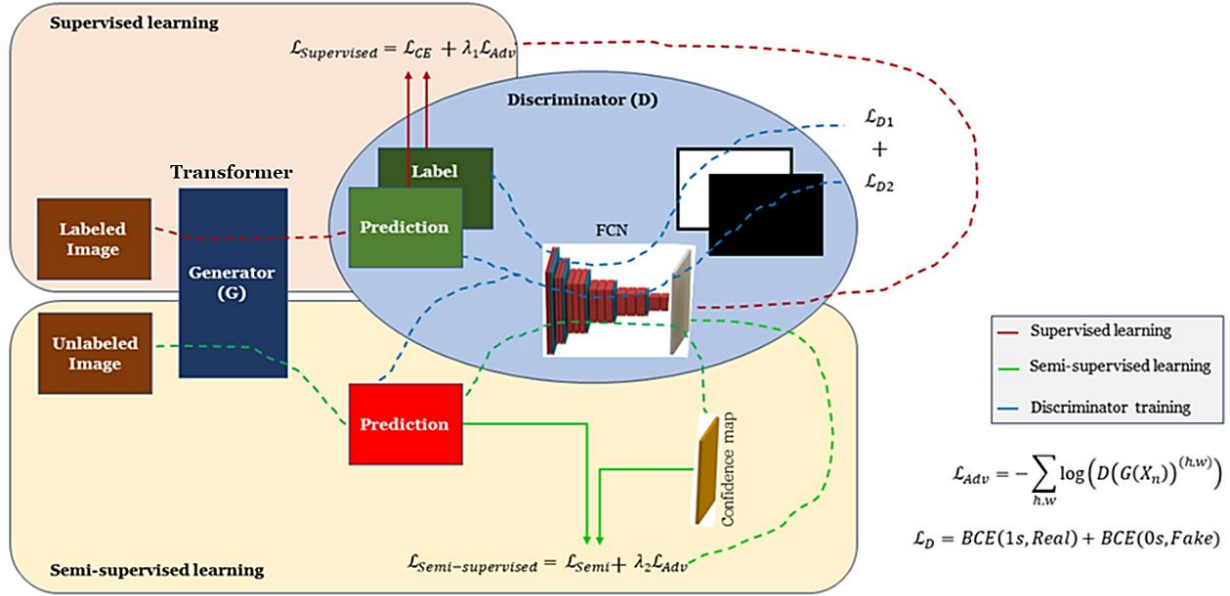

**Figure 7** Generative adversarial networks (GAN)-based segmentation model.

### 6.1.1 Generative Adversarial Network (GAN)

We prepare a GAN-based semi-supervised learning model as shown in **Figure 7**. It has two phases – supervised learning and semi-supervised learning. In supervised learning, we train the generator with labeled images. The generator ($G$) is a Mix Transformer (MiT-b3)-based segmentation model. A pair is then created by concatenating the labeled image and the prediction. A pair is also created by concatenating the ground truth and labeled image. These two pairs will be sent to the discriminator ($D$) to identify the fake and real pairs. The discriminator is a fully convolutional network (FCN) [10]. A segmentation loss function ($L_{seg}$) is used to train the generator and a discriminator loss function ($L_D$) to train the discriminator. In the semi-supervised learning phase, we train the generator with unlabeled data. This time we freeze the training of the discriminator, rather we generate a confidence map using it. This confidence map and the generator prediction are used to calculate the semi-supervised loss. Loss functions are defined as:

$$\mathcal{L}_{Adv} = -\sum_{h,w} \log\left(D(G(X_n))^{(h,w)}\right) \tag{11}$$

$$\mathcal{L}_D = BCE(1s, Real) + BCE(0s, Fake) \tag{12}$$

$$\mathcal{L}_{Supervised} = \mathcal{L}_{CE} + \lambda_1 \mathcal{L}_{Adv} \tag{13}$$

$$\mathcal{L}_{Semi-supervised} = \mathcal{L}_{Semi} + \lambda_2 \mathcal{L}_{Adv} \tag{14}$$

$$\mathcal{L}_{Semi} = -\sum_{h,w} \sum_{c \in C} I\left(D(S(X_n))^{(h,w)} > T_{semi}\right) \cdot \hat{y}_n^{(h,w,c)} \log\left(G(X_n)^{(h,w,c)}\right) \tag{15}$$

where $I(\cdot)$ is the indicator function.

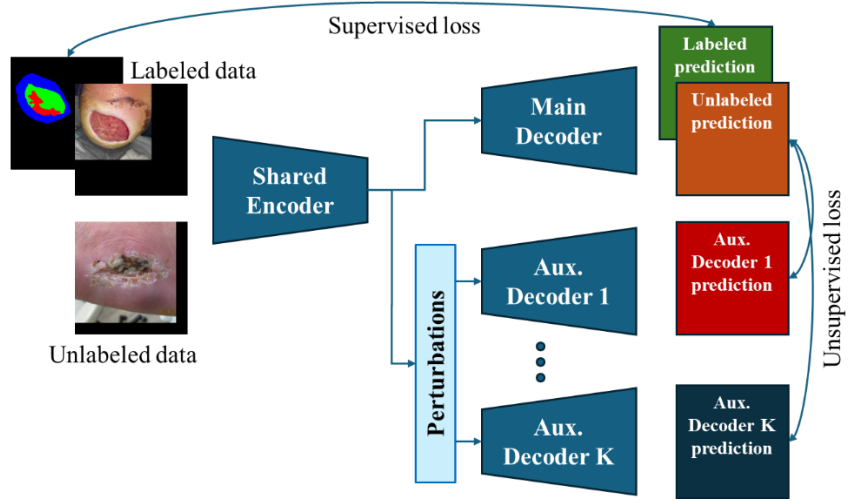

**Figure 8** Cross-consistency training (CCT)-based segmentation model.

### 6.1.2 Cross-Consistency Training (CCT)

**Figure 8** depicts the cross-consistency training (CCT) we used as a baseline, originally proposed by [14]. A labeled image and its annotation, along with an unlabeled image, are processed by a shared encoder and main decoder. This setup generates two predictions: one for the labeled data and another for the unlabeled data. A supervised loss is calculated using cross-entropy loss for the labeled predictions. Perturbations are applied to the main encoder output from the unlabeled data, and these perturbed versions pass through auxiliary decoders. An unsupervised loss, measured by mean squared error (MSE), is computed between the auxiliary predictions and the main decoder predictions for the unlabeled data. ResNet50 [46] serves as the encoder backbone. The study employs various perturbation functions, including transformation-based, prediction-based, feature-based, and random-based methods. Additional perturbation functions, such as random rotation, zoom, blurring, normalization, and elastic transformation, were also used alongside the seven mentioned in [14].

**Table 3** Supervised learning results for the DFUTissue dataset.

| Model | IoU | Precision | Recall | DSC |
| --- | --- | --- | --- | --- |
| SegFormer-b3 | 67.31 | 85.49 | 75.99 | 80.46 |
| U-Net | 56.31 | 84.21 | 62.96 | 72.05 |
| DeepLabV3+ | 59.0 | 86.51 | 65.44 | 75.13 |
| Ours | **73.75** | **87.69** | **82.27** | **84.89** |

**Table 4** Segmentation results for semi-supervised learning.

| Model | IoU | Precision | Recall | DSC |
|---|---|---|---|---|
| CCT | 67.41 | 81.41 | 79.68 | 80.53 |
| GAN | 70.96 | 81.70 | 84.36 | 83.01 |
| Ours | **77.99** | **88.14** | **87.14** | **87.64** |

## 6.2 Quantitative Analysis

### 6.2.1 Quantitative Analysis of The DFUTissue Dataset for Supervised Learning (SL) Phase

As outlined in the baseline, we conduct a comprehensive comparison of our model against three prominent models in the field: SegFormer-b3, U-Net, and DeepLabV3+. The results of this comparison are tabulated in **Table 3**. Notably, the dice scores for both U-Net and DeepLabV3+ were observed to fall below the 75% threshold. In contrast, SegFormer-b3 exhibited a more commendable performance, achieving a dice score of 80.46. However, our hybrid supervised model surpasses all three benchmark models, showcasing superior performance with a dice score of 84.89%. This outcome underscores the potency and efficacy of our approach, positioning it as a frontrunner for the semi-supervised stage.

### 6.2.2 Quantitative Analysis of The DFUTissue Dataset for Semi-Supervised Learning (SSL) Phase

As noted in the baseline, we conduct a comparison of our semi-supervised learning (SSL) approach with two widely used SSL methods – Generative Adversarial Networks (GAN)-based, and Cross-Consistency Training (CCT)-based. As shown in **Table 4**, GAN and CCT achieved an overall Dice Similarity Coefficient (DSC) of 80.53% and 83.01%, respectively, while our approach outperforms them with a DSC of 87.64%. Additionally, as shown in **Table 5**, we perform a label-wise comparison, generating metrics for individual tissues. CCT achieved DSC values of 53.84%, 89.45%, and 70.64% for fibrin, granulation, and callus, respectively. GAN exhibited slightly better performance with DSC values of 60.04%, 92.66%, and 70.95% for the same tissue sequence. Our model, however, surpassed both GAN and CCT for all three tissues, achieving DSC values of 69.01%, 94.11%, and 78.27%. Our proposed model demonstrated approximately 15% and 9% higher DSC in fibrin compared to GAN and CCT, respectively. Similarly, it achieved 4.66% and 1.45% higher DSC in granulation, and 7.5% and 7.32% in callus, respectively. It's noteworthy that fibrin and callus pose greater challenges in detection due to their intensity and appearance. Our model exhibited a minimum of a 7% increase in DSC for these two tissues compared to GAN and CCT.

**Table 5** Segmentation results for individual tissues for semi-supervised learning.

| Models | Metrics | Fibrin | Granulation | Callus |
|---|---|---|---|---|
| CCT | IoU | 36.83 | 80.92 | 54.60 |
| | Precision | 58.40 | 91.67 | 68.86 |
| | Recall | 49.94 | 87.34 | 72.51 |
| | DSC | 53.84 | 89.45 | 70.64 |
| GAN | IoU | 42.89 | 86.32 | 54.97 |
| | Precision | 52.29 | 90.71 | 73.02 |
| | Recall | 70.47 | 94.69 | 68.99 |
| | DSC | 60.04 | 92.66 | 70.95 |
| Ours | IoU | 52.68 | 88.87 | 64.29 |
| | Precision | 71.48 | 93.12 | 80.65 |
| | Recall | 66.70 | 95.11 | 76.02 |
| | DSC | 69.01 | 94.11 | 78.27 |

**Table 6** Validation results obtained by SemanticGAN [13].

| | Fibrin | Granulation | Callus |
|---|---|---|---|
| **mIoU** | 27.07 | 36.57 | 48.15 |

In addition, we attempted to use the semanticGAN [13], another GAN-based segmentation model. SemanticGAN is built on top of StyleGAN2 [47]. As shown in **Table 6**, we encounter challenges in achieving satisfactory validation performance for the DFUTissue dataset, attributed to two main reasons. Firstly, the StyleGAN2 of SemanticGAN is not pretrained on wound images. Moreover, StyleGAN2's synthesis quality is constrained to unimodal data, such as facial images, and is less effective for highly complex data or images with subtle details. Another contributing factor is that SemanticGAN necessitates a substantial amount of unlabeled data, which we currently lack.

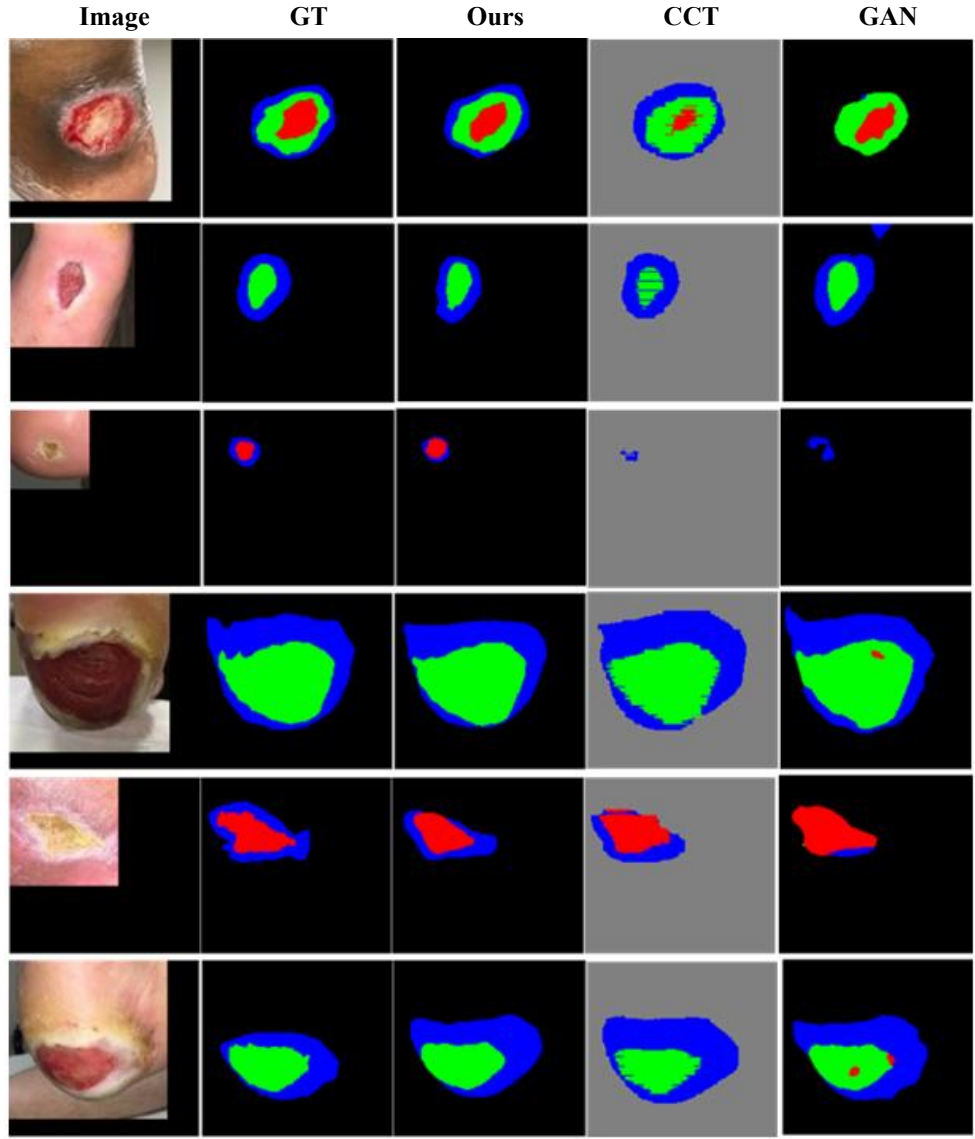

**Figure 9** Segmentation results for wound tissue dataset. Red, green, and blue are used to indicate fibrin, granulation, and callus, respectively. Images are cropped for better visualization.

### 6.3 Qualitative Analysis

**Figure 9** illustrates the results obtained from the DFUTissue dataset, with different colors representing distinct tissue types: red for fibrin, green for granulation, and blue for callus. In our comparative analysis, we evaluated our model against two semi-supervised learning (SSL) models – the generative adversarial networks (GAN) based model and the cross-consistency training (CCT) model. As depicted in **Figure 9**, our model exhibits superior performance compared to both the GAN-based and CCT models. It is observed that the GAN model erroneously predicted the presence of fibrin tissues, even though no such tissue actually exists. In contrast, among the three models, CCT demonstrates comparatively lower performance by generating unsmooth edges. Also, regarding the image in the third row of **Figure 9**, which had low resolution before zero-padding, both CCT and GAN failed to segment the tissues. However, our model performed significantly better than CCT and GAN in predicting them.

Table 7 Different decoder configurations.

| Decoder | Precision | Recall | DSC |
|---|---|---|---|
| No SE | 87.95 | 78.69 | 83.06 |
| With scSE | 86.04 | 81.81 | 83.87 |
| Ours | 87.69 | 82.27 | 84.89 |

Table 8 Performance analysis for different numbers of random selection.

| Random pick | Round | #Images | Epoch | Precision | Recall | DSC | Time (Hr.) |
|---|---|---|---|---|---|---|---|
| 25 | 1 | 103 | 493 | 87.52 | 84.15 | 85.80 | 1.32 |
|  | 2 | 128 | 372 | 89.57 | 82.95 | 86.14 | 0.95 |
|  | 3 | 153 | 125 | 89.14 | 84.45 | 86.73 | 0.63 |
|  | 4 | 178 | 384 | 86.09 | 87.80 | 86.94 | 1.19 |
| 50 | 1 | 128 | 479 | 86.90 | 84.18 | 85.52 | 1.73 |
|  | 2 | 178 | 437 | 88.38 | 83.77 | 86.01 | 2.75 |
|  | 3 | 228 | 488 | 88.05 | 86.11 | 87.07 | 3.19 |
|  | 4 | 278 | 433 | 88.14 | 87.14 | 87.64 | 3.92 |
| 100 | 1 | 178 | 85 | 89.16 | 83.48 | 86.23 | 0.52 |
|  | 2 | 278 | 499 | 89.14 | 83.72 | 86.35 | 4.21 |
|  | 3 | 378 | 178 | 89.05 | 85.38 | 87.17 | 1.08 |

## 6.4 Ablation study

We investigate the decoder of our model by examining the squeezing-and-excitation (SE)-based attention in the decoder. As shown in **Table 7**, we compared our P-scSE-fused decoder against two architectures – one with no SE module and another with a spatial and channel SE (scSE) module [38]. When no SE module is used, we achieved a DSC of 83.06%. This value increases to 83.87% when a scSE module is added to each decoder stage. However, our decoder, which uses the P-scSE module, achieved a DSC of 84.89.

Next, we assess the model's performance by randomly selecting varying numbers of images during the semi-supervised phase. As explained in the algorithm, for each round, we train the model $K$ times. On each occasion, we select $n$ unlabeled images along with their corresponding pseudo-labels. The results obtained for 25, 50, and 100 randomly chosen images are tabulated in **Table 8**. The best Dice Similarity Coefficient (DSC) of 87.64% is achieved when 50 images are selected. Although the DSC for 100 images is 87.17%, it requires less time compared to selecting 50 images. However, all random selections completed the semi-supervised phase within 12 hours, which is reasonable in terms of training time. After four rounds, we did not observe any improvement in any of the selections.

## 6.5 Wound region segmentation

Although the main goal of our study is multi-class segmentation, we also analyze our hybrid model's performance in binary segmentation using the publicly available Chronic Wound dataset [15], which includes 1010 labeled images of diabetic foot ulcers (810 for training, and 200 for inference). Experiments are conducted in supervised learning. As illustrated in **Table 9**, our proposed hybrid model demonstrates notable results, achieving an IoU of 86.89%, a precision of 93.98%, a recall of 92.01%, and a DSC of 92.99%. Notably, while our model leverages the Mix Transformer (MiT) architecture, first introduced in SegFormer [35], it exhibits a significant performance advantage over the original SegFormer model. Furthermore, our model surpasses the previously reported best-performing FUSegNet model in terms of both DSC and IoU. These findings collectively underscore the efficacy of our supervised model in accurately delineating wound regions within diabetic foot ulcer images.

Table 9 Comparison of binary segmentation results obtained by the proposed supervised model for the chronic wound dataset with state-of-the-art methods.

| Model | IoU | Precision | Recall | DSC |
| --- | --- | --- | --- | --- |
| MobileNetV2+CCL | NA | 91.01 | 89.97 | 90.47 |
| LinkNet-EffB1 + UNet-EffB2 | 85.51 | 92.68 | 91.80 | 92.07 |
| DeepLabV3Plus | 85.19 | 92.75 | 91.27 | 92.00 |
| Swin-Unet | 79.30 | 89.94 | 87.02 | 88.46 |
| DDRNet | 57.64 | 80.86 | 66.75 | 73.13 |
| SegFormer-b5 | 83.58 | 92.21 | 89.94 | 91.06 |
| FUSegNet | 86.40 | **94.40** | 91.07 | 92.70 |
| Ours | **86.89** | 93.98 | **92.01** | **92.99** |

The qualitative results for the Chronic Wound dataset are illustrated in **Figure 10**. Outputs are presented for different skin colors, with images selected at various resolutions. The image resolutions from row-1 to row-5 in **Figure 10** are 70 × 138, 89 × 127, 82 × 142, 48 × 117, and 40 × 48, respectively. Additionally, wound regions exhibit different color intensities. For instance, the wound region in row-1 has a pinkish color, whereas it is considerably darker in row-4 and row-5. To enhance visualization, images are cropped, and wound regions are zoomed. As depicted in **Figure 10**, our supervised model outperformed the other models in predicting the wound regions. Green borders indicate predicted wound regions, while red borders represent the original wound regions.

## 6.6 Limitation

The semi-supervised learning (SSL) phase is an iterative method. If there is a large amount of unlabeled data, the number of rounds might increase, thereby increasing the training time. However, traditional SSL methods require a large amount of unlabeled data to generalize the model. On the other hand, our approach has been found effective even with a small unlabeled dataset. With only 600 unlabeled data, our proposed framework exhibited superior performance compared to the CCT and GAN models.

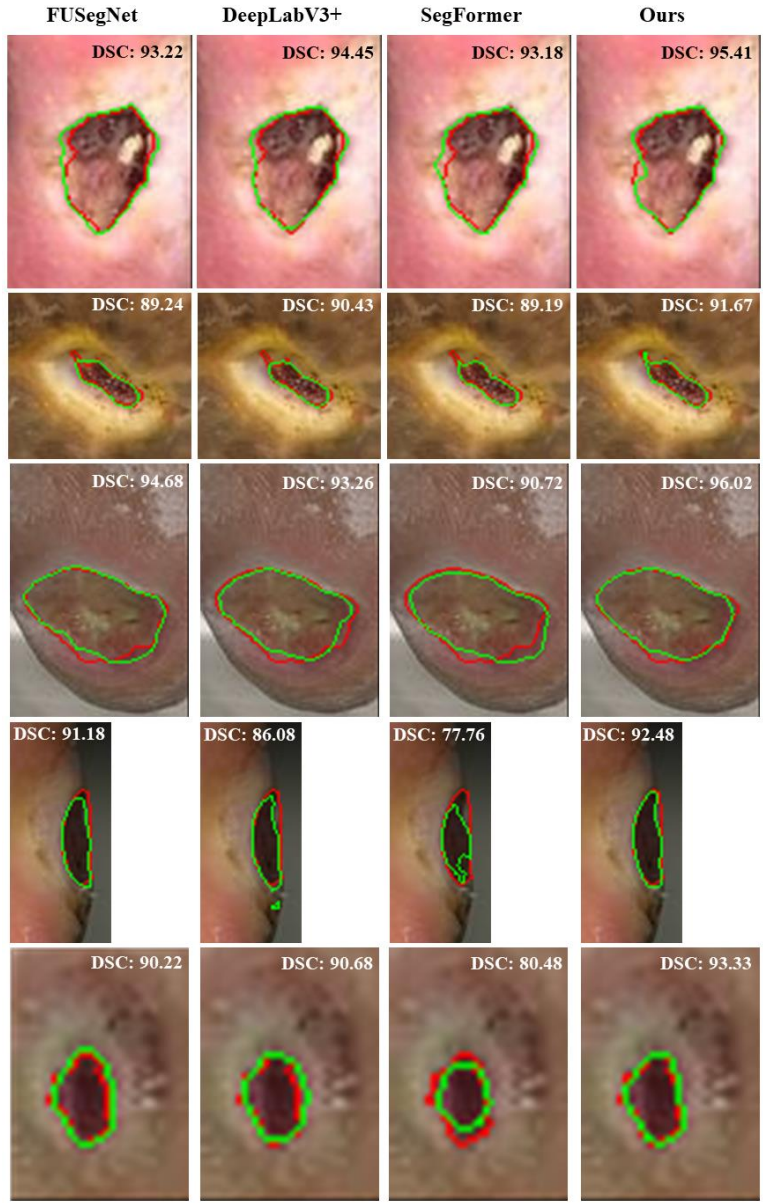

**Figure 10** Segmentation results for chronic wound dataset. The original and predicted boundaries are shown in red and green, respectively. Images are cropped for better visualization.

## 7 Conclusion

Our research represents a pioneering effort in the relatively underexplored field of wound tissue segmentation in DFUs, which has seen limited dedicated works. The main reason is the lack of an annotated dataset. To address this gap, we release the DFUTissue dataset, which contains 110 annotated DFU images. We believe that the DFUTissue dataset will encourage the research community to contribute to wound tissue segmentation. We also perform a pilot study to tackle the challenges of segmenting wound tissues, which requires a deep understanding of tissue characteristics, healing dynamics, and morphological variations. By addressing these challenges and offering innovative solutions, our work makes a significant contribution to this emerging field. We aim to expand the reach of our work by generating more labeled wound tissue data and optimizing our model. This includes investigating different stopping criteria, such as combining validation loss and metrics, to further enhance our model's effectiveness and establish it as a reference for evaluating wound segmentation algorithms.

# 8 Acknowledgements

The present work is partially supported by the Office of Advanced Cyberinfrastructure of National Science Foundation under Grant No. 2232824.